\input{psfig.sty}

\documentclass[12pt,preprint]{aastex}

\begin{document}

\title{Chandra Observation  of the  Persistent Emission from  the Dipping
  Source XB 1916--053 }

\author{R. Iaria\altaffilmark{1},  
   T. Di   Salvo\altaffilmark{1}, 
   G. Lavagetto\altaffilmark{1}, 
   N. R. Robba\altaffilmark{1},
   L. Burderi\altaffilmark{2} }

\altaffiltext{1}{Dipartimento di Scienze Fisiche ed Astronomiche,
Universit\`a di Palermo, via Archirafi 36 - 90123 Palermo, Italy; email:iaria@fisica.unipa.it}
\altaffiltext{2}{Universit\`a degli Studi di Cagliari, Dipartimento
di Fisica, SP Monserrato-Sestu, KM 0.7, 09042 Monserrato, Italy}
%\altaffiltext{2}{Universit\`a degli Studi di Cagliari, Dipartimento
%di Fisica, SP Monserrato-Sestu, KM 0.7, 09042 Monserrato, Italy}

\begin{abstract}
  
  We present  the results of a  50 ks long Chandra  observation of the
  dipping  source  XB 1916--053.   During  the  observation two  X-ray
  bursts  occurred and  the  dips  were not  present  at each  orbital
  period.   From the zero-order  image we  estimate the  precise X-ray
  coordinates of the source with a 90\% uncertainty of 0.6\arcsec.  In
  this  work we  focus on  the spectral  study of  discrete absorption
  features,  during the  persistent  emission, using  the High  Energy
  Transmission  Grating Spectrometer on  board the  Chandra satellite.
  We  detect,  for the  first  time in  the  1st-order  spectra of  XB
  1916--053,    absorption   lines    associated    to   \ion{Ne}{10},
  \ion{Mg}{12}, \ion{Si}{14} and \ion{S}{16}, and confirm the presence
  of the \ion{Fe}{25} and  \ion{Fe}{26} absorption lines with a larger
  accuracy  with respect  to  the previous  XMM  EPIC pn  observation.
  Assuming  that  the line  widths  are  due to  a  bulk  motion or  a
  turbulence associated to the  coronal activity, we estimate that the
  lines  are produced  in  a photoionized  absorber  distant from  the
  neutron star $4 \times 10^{10}$ cm, near the disk edge.

\end{abstract}

\maketitle

\section{Introduction}

Low Mass  X-ray Binaries (LMXBs) consist  of a low mass  star ($\le 1$
M$_\odot$)  and a  neutron star  (NS), generally  with a  weak magnetic
field  ($B \le  10^{10}$  G). In  these  systems the  X-ray source  is
powered by  the accretion  of mass overflowing  the Roche lobe  of the
companion star and forming an  accretion disk around the NS. Different
inclinations of  the line of sight  with respect to  the orbital plane
can  explain the  different  properties of  the  lightcurves of  these
systems. At low  inclinations ($i \le 70^\circ$) eclipse  and dips are
not observed  in the  lightcurves, while they  can be present  at high
inclination  angles  because  the  companion  star  and/or  the  outer
accretion disk intercept  the line of sight. 

About  10  LMXBs  are known  to  show  periodic  dips in  their  X-ray
lightcurves. The  dips recur at the  orbital period of  the system and
are  probably caused  by a  thicker  region in  the outer  rim of  the
accretion disk, formed  by the impact with the disk  of the gas stream
from the companion star (White  \& Swank, 1982).  The dip intensities,
lengths and shapes change from  source to source, from cycle to cycle.
For systems seen  almost edge-on, X-ray emission is  still visible due
to the  presence of an extended  accretion disk corona  (ADC, White \&
Holt 1982). The ADC can be periodically eclipsed by the companion star
and have  a radius between $10^9$  and $5 \times 10^{10}$  cm, with an
appreciable fraction  (from 10\%  up to 50  \%) of the  accretion disk
radius  (see   Church  \&  Balucinska-Church,  2001   for  a  review).
Information  on  the emitting  region  in  LMXBs  can be  obtained  by
studying  the  spectrum of  these  high  inclination  sources and  its
evolution during the dips.  The energy spectrum of the dipping sources
can  be  well  described   using  two  different  scenarios.   In  the
``absorbed plus unabsorbed'' scenario  (e.g., Parmar et al.  1986) the
persistent (non-dipping) spectral shape  was used to model the spectra
during the dips,  but is divided into two parts.   One part is allowed
to be absorbed, whereas the  other one is not.  The spectral evolution
during  dipping is well  modelled by  a large  increase in  the column
density of the absorbed component, and a decrease of the normalization
of  the  unabsorbed  component.   The latter  component  is  generally
attributed   to  electron   scattering  in   the  absorber.    In  the
``progressive covering''  scenario (e.g., Church  \& Balucinska-Church
1995), the  X-ray emission is  assumed to originate from  a point-like
blackbody, or disk-blackbody component (probably from the neutron star
surface or  the inner region of  the accretion disc),  together with a
power law (probably from the  extended ADC).  These two components can
be used  to fit the spectrum  both during the  persistent emission and
the  dipping activity.  Moreover,  during the  dipping the  model well
describes  the spectral  changes due  to the  partial  and progressive
covering  of the  power-law  emission from  an  extended source.   The
absorption   of  the   point-like   component  is   allowed  to   vary
independently  from that  of the  extended component,  and  usually no
partial covering is included because during the dipping activity it is
fully  covered.   Both  these  approaches  have  been  applied  to  XB
1916--053  (e.g.,   Yoshida  et  al.   1995;  Church   et  al.   1997,
respectively).

The  improved  sensitivity  and  spectral resolution  of  Chandra  and
XMM-Newton are  allowing to  observe narrow absorption  features, from
highly  ionized ions  (H-like and  He-like),  in a  larger and  larger
number  of X-ray  binaries.   These features  were  detected from  the
micro-quasars GRO J1655--40 (Ueda et  al.  1998; Yamaoka et al.  2001)
and GRS  1915+105 (Kotani et al.   2000; Lee et  al.  2002).  Recently
the  Chandra  High-Energy  Transmission Grating  Spectrometer  (HETGS)
observations of  the black hole  candidate H 1743--322 (Miller  et al.
2004)  have revealed  the  presence of  blue-shifted \ion{Fe}{25}  and
\ion{Fe}{26}  absorption   features  suggesting  the   presence  of  a
highly-ionized  outflow.    All  LMXBs  those   exhibit  narrow  X-ray
absorption  features are  all known  dipping sources  (see Table  5 of
Boirin  et al.   2004)  except for  GX  13+1. This  source shows  deep
blue-shifted  Fe  absorption features  in  its  HETGS spectrum,  again
indicative of outflowing material (Ueda et al. 2004).

More recent Diaz Trigo et al.  (2005) has modelled the changes in both
the X-ray continuum  and the Fe absorption features  during dipping of
six bright  LMXBs  observed by XMM-Newton.  They concluded that
the  dips are  produced by  an increase  in the  column density  and a
decrease  in the  ionization  state of  the  highly ionized  absorber.
Moreover, outside of  the dips, the absorption line  properties do not
vary strongly with orbital phase, this implies that the ionized plasma
has a cylindrical geometry with  a maximum column density close to the
plane of the  accretion disk.  Since dipping sources  are normal LMXBs
viewed  from close  to the  orbital  plane this  implies that  ionized
plasmas are a common feature  of LMXBs.  Similar results were obtained
by  Boirin et  al.  (2005)  studying  a XMM-Newton  observation of  XB
1323--619.

XB  1916--053 (4U  1916--05) is  a  dipping source  with the  shortest
period of all dipping sources of  50 min (Walter et al., 1982), and it
is also  notable because of the  difference of 1 \%  between the X-ray
and  optical  periods  (see  Callanan  et  al.,  1995  and  references
therein).  Recently  Retter et al.  (2002) have  favored the superhump
model  to explain  this discrepancy.   The superhump  model  invokes a
precessing  accretion  disk  which  identifies  the  X-ray  period  as
orbital.  XB 1916--053 was observed with OSO-8 and Ginga above 10 keV.
From the  OSO-8 results  of White  \& Swank (1982)  it was  clear that
dipping persisted  up to  20 keV.  Using  BeppoSAX data Church  et al.
(1998) showed  that the spectrum  of the dipping source  extends above
100 keV.

Boirin et  al.  (2004)  studied a 17  ks XMM-Newton  observation using
EPIC pn  and RGS  data in  timing mode, the  exposure time  during the
persistent emission was  10 ks.  The authors detected,  in the EPIC pn
data,  a   \ion{Fe}{25}  K$_\alpha$  and   a  \ion{Fe}{26}  K$_\alpha$
absorption line  centered at 6.65 and  6.95 keV, with  upper limits on
the  line  widths of  100  and  140  eV, respectively;  moreover  they
marginally  detected,  in the  RGS  data between  0.5  and  2 keV,  an
absorption  line centered  at  1.48 keV  with  an upper  limit of  the
corresponding line width of 41  eV and, finally, an absorption edge at
0.99  keV.   The  absorption  line  of  1.48  keV  was  associated  to
\ion{Mg}{12}  K$_\alpha$ and  the  absorption edge  was associated  to
moderately  ionized  Ne  and/or  Fe.   Using  the  ratio  between  the
\ion{Fe}{25}  and \ion{Fe}{26}  column density  they have  estimated a
ionization  parameter log($\xi$)  of  3.92 erg  cm  s$^{-1}$.  From  a
combined  analysis  during  and  out  of the  dipping  intervals  they
concluded that during the dipping activity the absorber is composed by
cooler material.

In this work we present a spectral analysis of the persistent emission
from XB 1916--053  in the 0.8--10 keV energy range using  a 50 ks long
Chandra  observation.   The observation  covered  entirely 16  orbital
period however  we noted that  the dips were  not present at  all.  We
clearly detected,  for the first time  in the spectra  of this source,
the presence of  the \ion{Ne}{10} K$_\alpha$, \ion{Mg}{12} K$_\alpha$,
\ion{Si}{14} K$_\alpha$,  and \ion{S}{16} K$_\alpha$  absorption lines
and  confirmed  the  presence  of  the  \ion{Fe}{25}  K$_\alpha$,  and
\ion{Fe}{26} K$_\alpha$  absorption lines, although  the better energy
resolution  of  Chandra and  the  larger  statistics  allowed to  well
determine the  widths of  each line.  We  discuss that  the absorption
lines were produced  in a photoionized absorber placed  at the edge of
the accretion disk, probably the same absorber producing the dips when
it is less photoionized.

\section{Observation} 

XB 1916--053 was observed with  the Chandra observatory on 2004 Aug 07
from 02:27:22 to  16:07:31 UT using the HETGS.   The observation had a
total integration  time of  50 ks, and  was performed in  timed graded
mode.  The HETGS  consists of two types of  transmission gratings, the
Medium Energy  Grating (MEG) and  the High Energy Grating  (HEG).  The
HETGS  affords  high-resolution  spectroscopy  from  1.2  to  31  \AA\
(0.4--10  keV)  with a  peak  spectral  resolution of  $\lambda/\Delta
\lambda  \sim 1000$ at  12 \AA\  for HEG  first order.   The dispersed
spectra  were recorded  with an  array of  six  charge-coupled devices
(CCDs)  which are  part  of the  Advanced  CCD Imaging  Spectrometer-S
(Garmire             et             al.,            2003)\footnote{See
  http://asc.harvard.edu/cdo/about\_chandra  for  more details.}.   We
processed the  event list using available software  (FTOOLS v6.0.2 and
CIAO v3.2  packages) and  computed aspect-corrected exposure  maps for
each spectrum, allowing  us to correct for effects  from the effective
area of the CCD spectrometer.

The brightness  of the source required additional  efforts to mitigate
``photon pileup'' effects. A 512  row ``subarray'' (with the first row
= 1) was applied during the observation reducing the CCD frame time to
1.7  s.  Pileup distorts  the count  spectrum because  detected events
overlap  and  their  deposited  charges  are  collected  into  single,
apparently more  energetic, events.   Moreover, many events  ($\sim 90
\%$) are  lost as the grades of  the piled up events  overlap those of
highly energetic background particles and  are thus rejected by the on
board software.  We, therefore, ignored the zeroth-order events in our
spectral analysis.  On  the other hand, the grating  spectra were not,
or only  moderately (less  than 10 \%),  affected by pileup.   In this
work we  analysed the 1st-order HEG  and MEG spectra; since  a 512 row
subarray  was applied  the 1st-order  HEG  and MEG  energy range  were
shrinked to 1--10 keV and 0.8--7 keV, respectively.

\begin{figure}
\resizebox{\hsize}{!}{\includegraphics{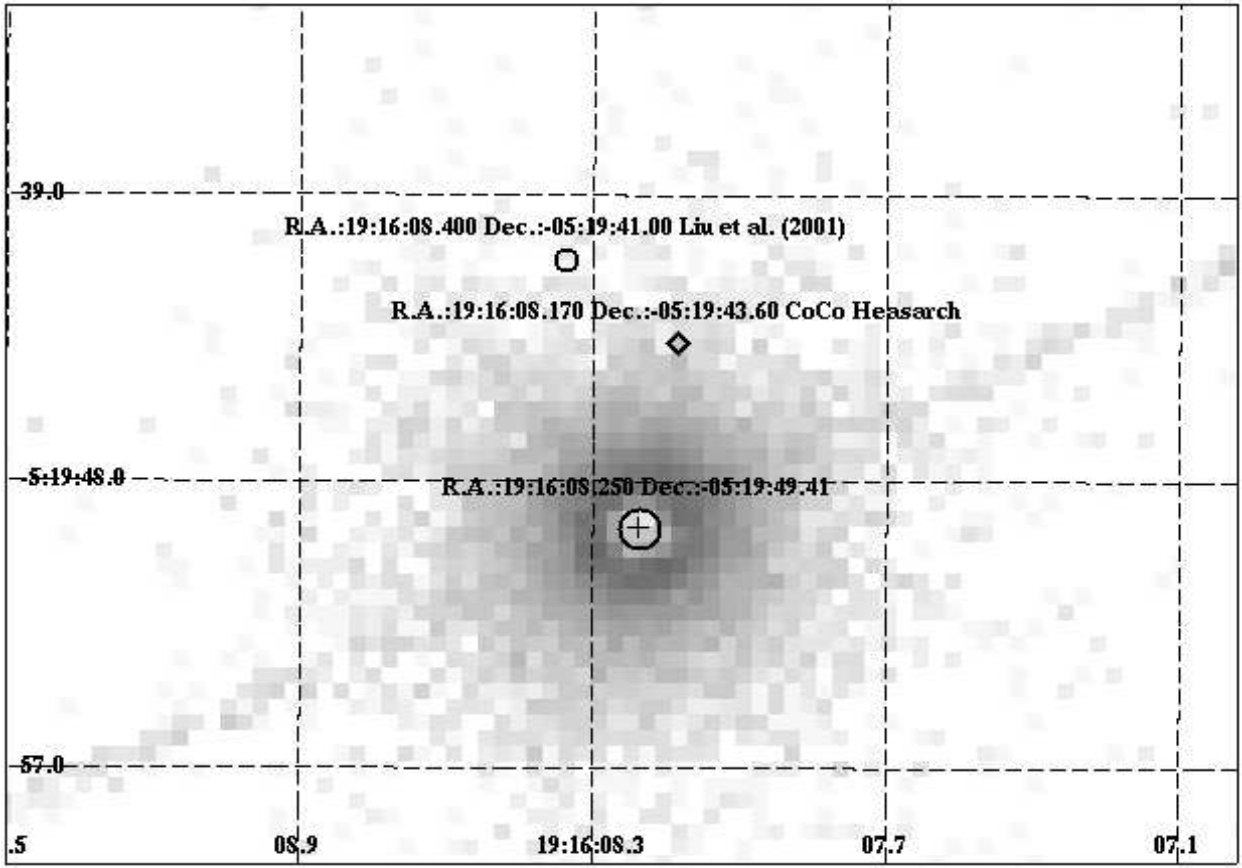}}
%\plotone{f1.ps}
\caption{Region  of sky  around  the Chandra  zero-order  image of  XB
  1916--053. The  coordinate system is referred to  B1950.  The circle
  centered  to  the  best  estimation  (cross point)  of  the  Chandra
  position has a radius of 0.6\arcsec. The diamond point indicates the
  position  of   XB  1916--053   (B1950)  reported  by   the  Heasarch
  ``Coordinate Converter'' tool (see text), the circle point indicates
  the  position of  the optical  counterpart of  XB  1916--053 (B1950)
  reported by Liu  et al.  (2001). The angular  separation between the
  Chandra  position  and the  position  reported  by the  ``Coordinate
  Converter'' tool is 6\arcsec.}
\label{fig0}
\end{figure}

To  determine the  zero-point position  in the  image as  precisely as
possible,  we estimated the  mean crossing  point of  the zeroth-order
readout trace  and the tracks  of the dispersed  HEG and MEG  arms. We
obtained    the   following    coordinates:   R.A.=$19^h18^m47^s.871$,
Dec.=$-05^{\circ}  14\arcmin  17\arcsec.09$   (J2000.0,  with  a  90\%
uncertainty circle of the absolute position of 0.6\arcsec\footnote{See
  http://cxc.harvard.edu/cal/ASPECT/celmon/  for more  details.}).  We
compared the Chandra position of XB 1916--053 referred to B1950 to the
coordinates  of the  optical counterpart  previously reported  (Liu et
al.,  2001 and references  therein) finding  an angular  separation of
8.7\arcsec.  Moreover  we considered  the coordinates of  XB 1916--053
reported    by    the   the    NASA    HEASARCH   tool    ``Coordinate
Converter''\footnote{See
  http://heasarch.gsfc.nasa.gov/cgi-bin/Tools/convcoord/convcoord.pl}
and compared those  to the Chandra position achieved  by ourself.  The
coordinates  obtained  with  the  tool were  R.A.=  $19^h18^m47^s.78$,
Dec.=-05$^{\circ}$14\arcmin 11\arcsec.2  (referred to J2000.0), having
an  angular  separation  from  the Chandra  coordinates  of  6\arcsec.
Unfortunately it was  not possible a comparison with  the previous XMM
observation because, in that case, the data of XB 1916--053 were taken
by all the EPIC cameras in  timing mode (see Boirin et al., 2004).  In
Fig.~\ref{fig0}  we  reported  a  region  of sky  around  the  Chandra
zero-order  image of  the  XB 1916--053  using  the coordinate  system
referred  to B1950  in  order  to compare  the  Chandra position,  the
``Coordinate  Converter''  tool  position,  and the  position  of  the
optical counterpart (Liu et al., 2001).

\begin{figure}
\resizebox{\hsize}{!}{\includegraphics{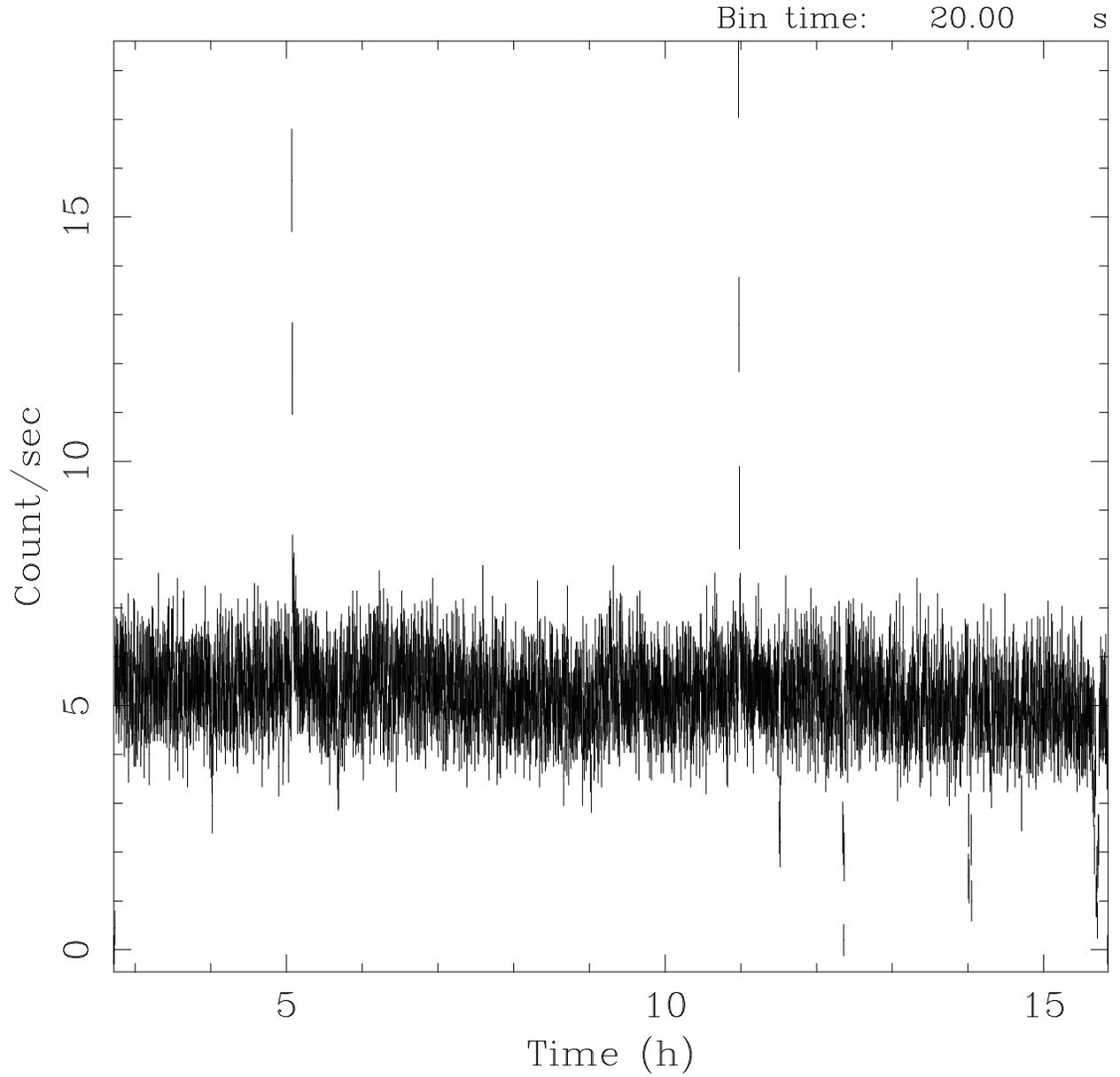}}
%\plotone{f1.ps}
\caption{The 50 ks lightcurve  of XB 1916--053. During the observation
  two  bursts   and  four  dips  were  observed.    The  used  events
  corresponds to the positive first-order  HEG.  The bin time is 20 s.}
\label{fig1}
\end{figure}

In Fig.~\ref{fig1} we showed the  20 s bin time lightcurve taking into
account only  the events  in the positive  first-order HEG.   The mean
count rate in  the persistent state was 6  count s$^{-1}$.  During the
observation two  bursts were observed,  the count rate at  their peaks
was  a  factor  three  larger  than during  the  persistent  emission.
Moreover we  observed only four  dips (see Fig.~\ref{fig1})  which did
not show a regular periodicity  as observed in the previous XMM-Newton
observation (Boirin et al., 2004);  in fact, while the second observed
dip occurred  $\sim 50$ min after  the first one, as  aspected for this
source,  the other  dips occurred  after temporal  intervals  two times
larger  ($\sim 1.7$  h),  indicating  that the  central  region of  XB
1916--053 was not occulted at every orbital passage.

\section{Spectral Analysis of the Persistent Emission}

We selected the  1st-order spectra from the HETGS  data, excluding the
bursts  and the dips,  with a  total exposure  time of  the persistent
emission  of 42.3  ks. Data  were  extracted from  regions around  the
grating arms; to avoid overlapping between HEG and MEG data, we used a
region size  of 25 and  33 pixels for  the HEG and  MEG, respectively,
along  the cross-dispersion  direction.  The  background  spectra were
computed, as usual,  by extracting data above and  below the dispersed
flux.  The contribution  from the background is $0.4  \%$ of the total
count  rate.  We  used  the  standard CIAO  tools  to create  detector
response files  (Davis 2001) for the HEG  -1 (MEG -1) and  HEG +1 (MEG
+1) order  (background-subtracted) spectra.  After  verifying that the
negative and  positive orders were  compatible with each other  in the
whole   energy  range   we  coadded   them  using   the   script  {\it
  add\_grating\_spectra} in the CIAO software, obtaining the 1st-order
MEG spectrum and the 1st-order  HEG spectrum.  Finally we rebinned the
resulting 1st-order  MEG and 1st-order  HEG spectra to 0.015  \AA\ and
0.0075  \AA, respectively.   It is  worth  to note  that the  absolute
wavelength accuracy, connected to  the error of 0.6\arcsec\ associated
to the  absolute source  position, is $\pm  0.006 {\rm \AA}$  and $\pm
0.011 {\rm \AA}$ for HEG and MEG, respectively.

\begin{figure}
\resizebox{\hsize}{!}{\includegraphics{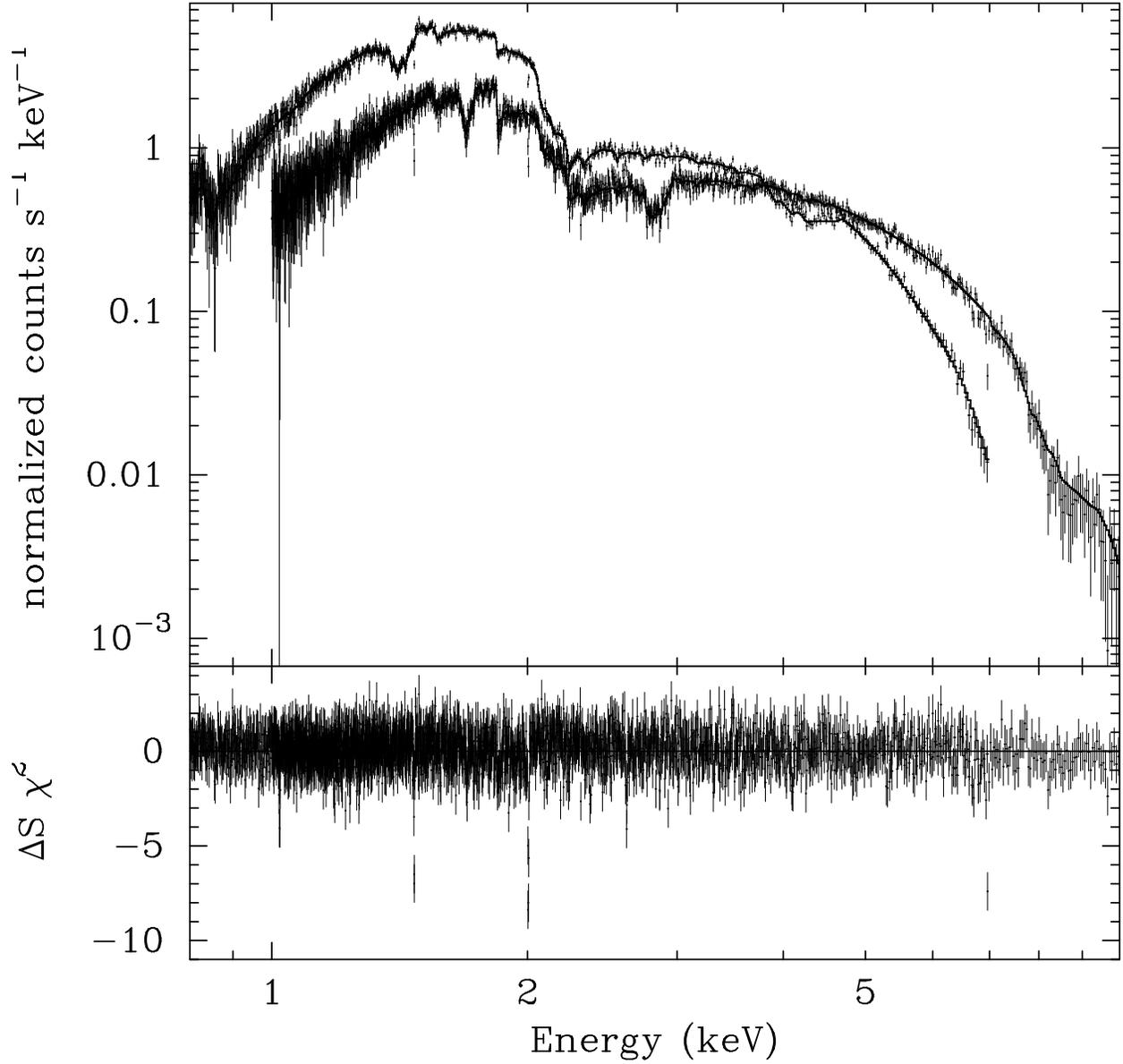}}
%\plotone{f1.ps}
\caption{Data and residuals of  the persistent emission in the energy
  range 0.8--10  keV for  the rebinned 1st-order  MEG and  HEG spectra
  (see text).  The continuum is fitted by an absorbed power law.  Five
  absorption features  are clearly  evident at 1,  1.5, 2, 2.5,  and 7
  keV, respectively. }
\label{fig2}
\end{figure}

\begin{figure}
\resizebox{\hsize}{!}{\includegraphics{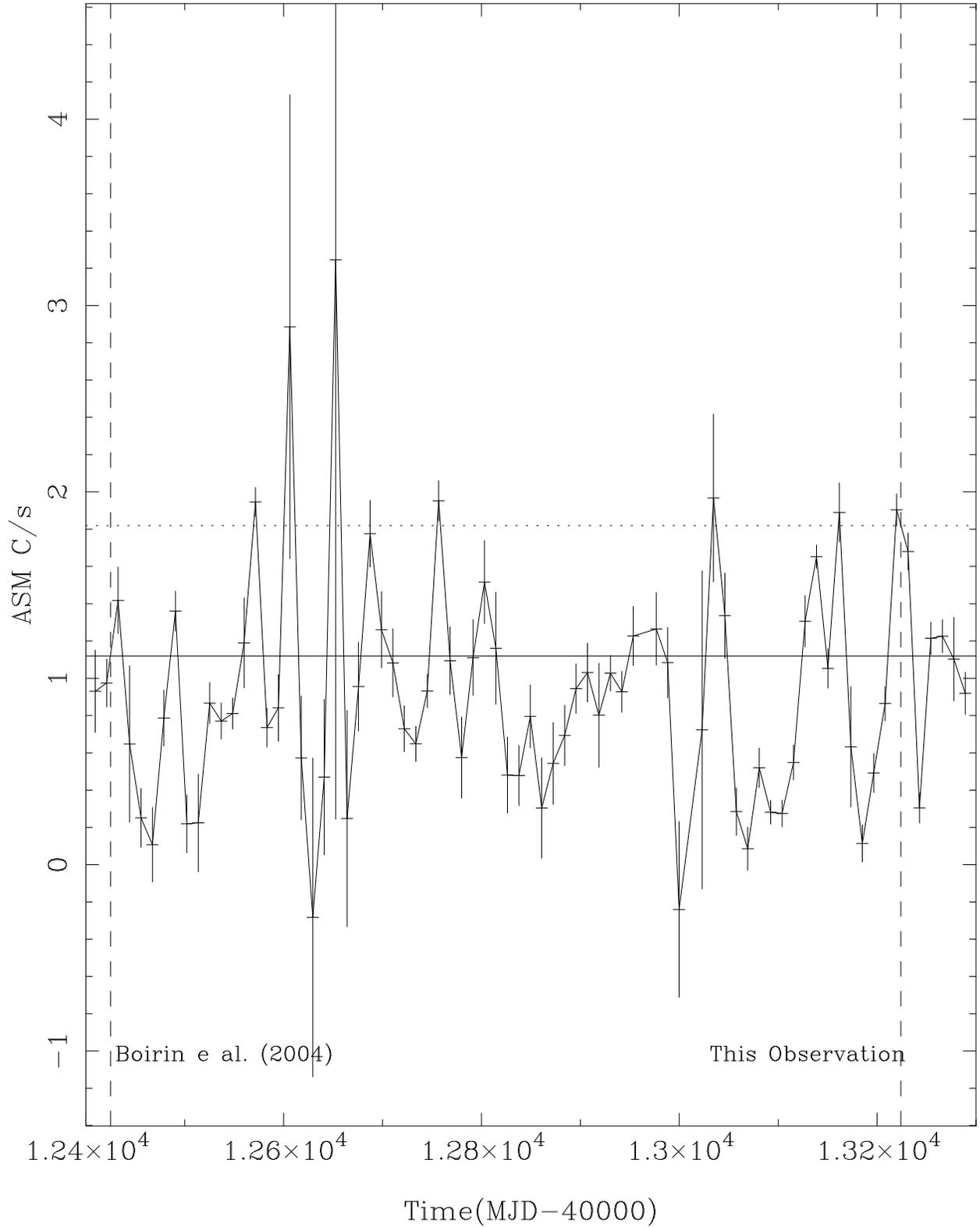}}
%\plotone{f1.ps}
\caption{The RXTE ASM lightcurve  of XB 1916--053. The dashed vertical
  lines indicate the start time of the previous XMM observation and of
  this observation. During the XMM  observation the ASM count rate was
  1.12 C/s  (solid horizontal line),  during this observation  the ASM
  count rate was 1.32 C/s (dotted horizontal line).}
\label{fig2b}
\end{figure}

To fit  the continuum we used  the rebinned spectra in  the 0.8--7 keV
and 1--10  keV for first-order MEG and  first-order HEG, respectively.
We well fitted  the continuum using an absorbed  power law and adding,
in the MEG data, a systematic  edge at around 2.07 keV with a negative
optical depth  of $\sim  -0.2$ to take  in account of  an instrumental
artifact  (see  Miller et  al.   2002,  and  references therein).   We
obtained a  $\chi^2$(d.o.f.)  of  2325(2395).  We found  an equivalent
hydrogen column  of N$_H  = 0.44 \times  10^{22}$ cm$^{-2}$,  a photon
index   of  1.5,   and  a   power-law  normalization   of   0.11.   In
Fig.~\ref{fig2} we reported the data and the residuals with respect to
the  continuum described  above.  The  presence of  several absorption
features  was clearly  evident  in the  residuals.  The  corresponding
absorbed flux and the unabsorbed luminosity assuming a distance to the
source  of 9.3  kpc were  $\sim  7.3 \times  10^{-10}$ ergs  cm$^{-2}$
s$^{-1}$ and  $\sim 7.5 \times  10^{36}$ ergs s$^{-1}$ in  the 0.6--10
keV,  respectively. We  noted that  the  luminosity was  a factor  1.7
larger than during the XMM observation  (see Tab.  1 in Boirin et al.,
2004).  To  have a  confirm of  this result we  analysed the  RXTE ASM
lightcurve. We observed that during  the XMM observation the ASM count
rate was 1.12 C/s, while  during the Chandra observation the ASM count
rate  was 1.82  C/s  with an  increase  of intensity  of almost  1.63,
similar  to  the  value  obtained  from  the  spectral  analysis.   In
Fig.~\ref{fig2b} we  reported the ASM lightcurve of  XB 1916--053, the
dashed vertical lines indicated the  start time of the XMM and Chandra
observations, the solid horizontal line indicated the level of the ASM
count  rate  during the  XMM  observation,  and,  finally, the  dotted
horizontal line indicated  the level of the ASM  count rate during the
Chandra observation.

\begin{figure}
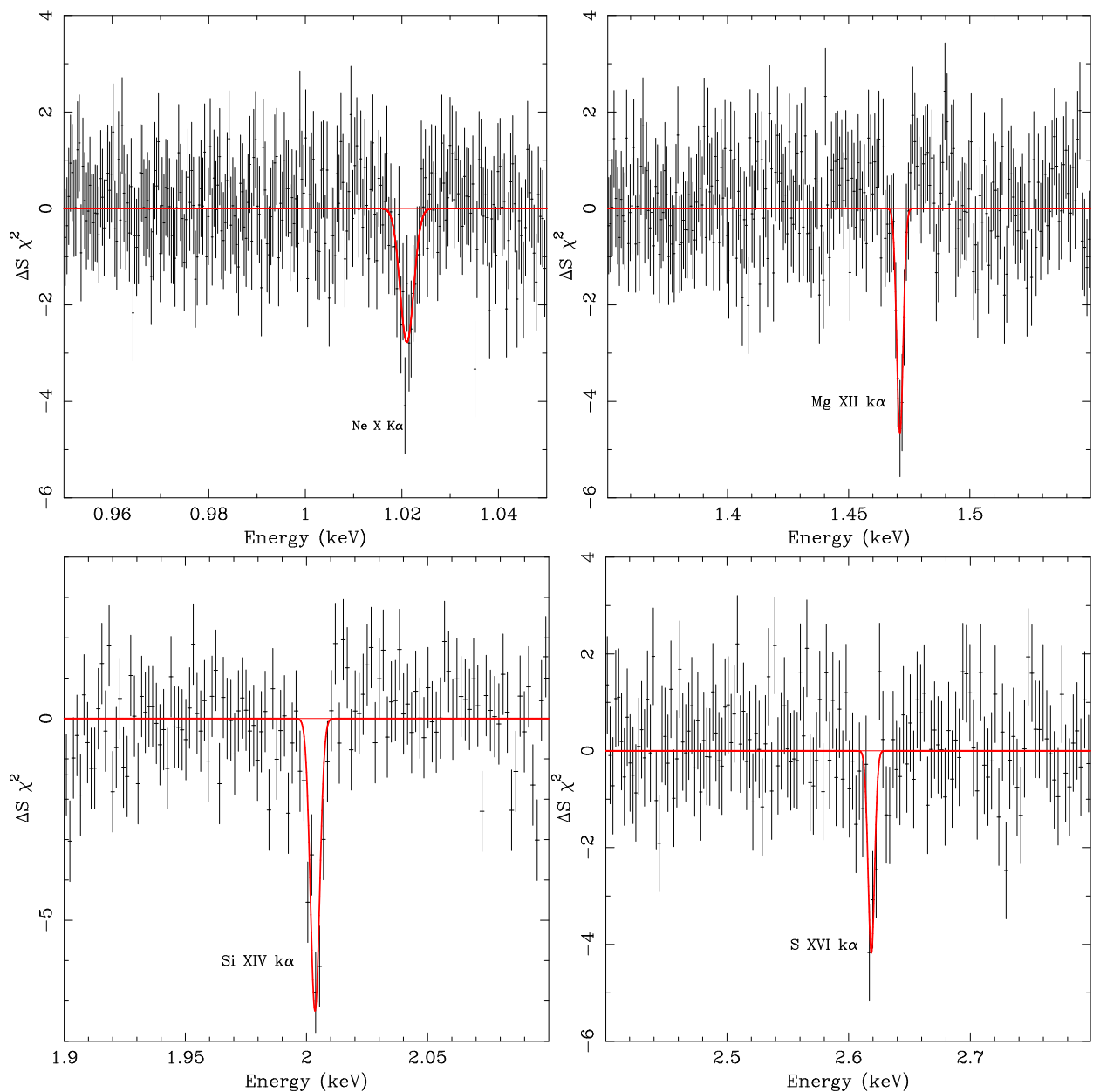

  \resizebox{\hsize}{!}{\includegraphics{f5a.eps}
    \includegraphics{f5b.eps}}\\
  \resizebox{\hsize}{!}{\includegraphics{f5c.eps}\includegraphics{f5d.eps}}
  \caption{Residuals of  the  1st-order  MEG spectrum with
    respect to  the best-fit model  of the continuum reported  in Tab.
    \ref{tab1}. In  the four panels  the residuals are plotted  in the
    energy  intervals around  the observed  absorption  features which
    were  identified  as   \ion{Ne}{10}  K$_\alpha$  (top-left  panel),
    \ion{Mg}{12} K$_\alpha$  (top-right panel), \ion{Si}{14} K$_\alpha$
    (bottom-left panel),  \ion{S}{16} K$_\alpha$ (bottom-right panel).}
\label{fig3}
\end{figure}

\begin{figure}
\resizebox{\hsize}{!}{\includegraphics{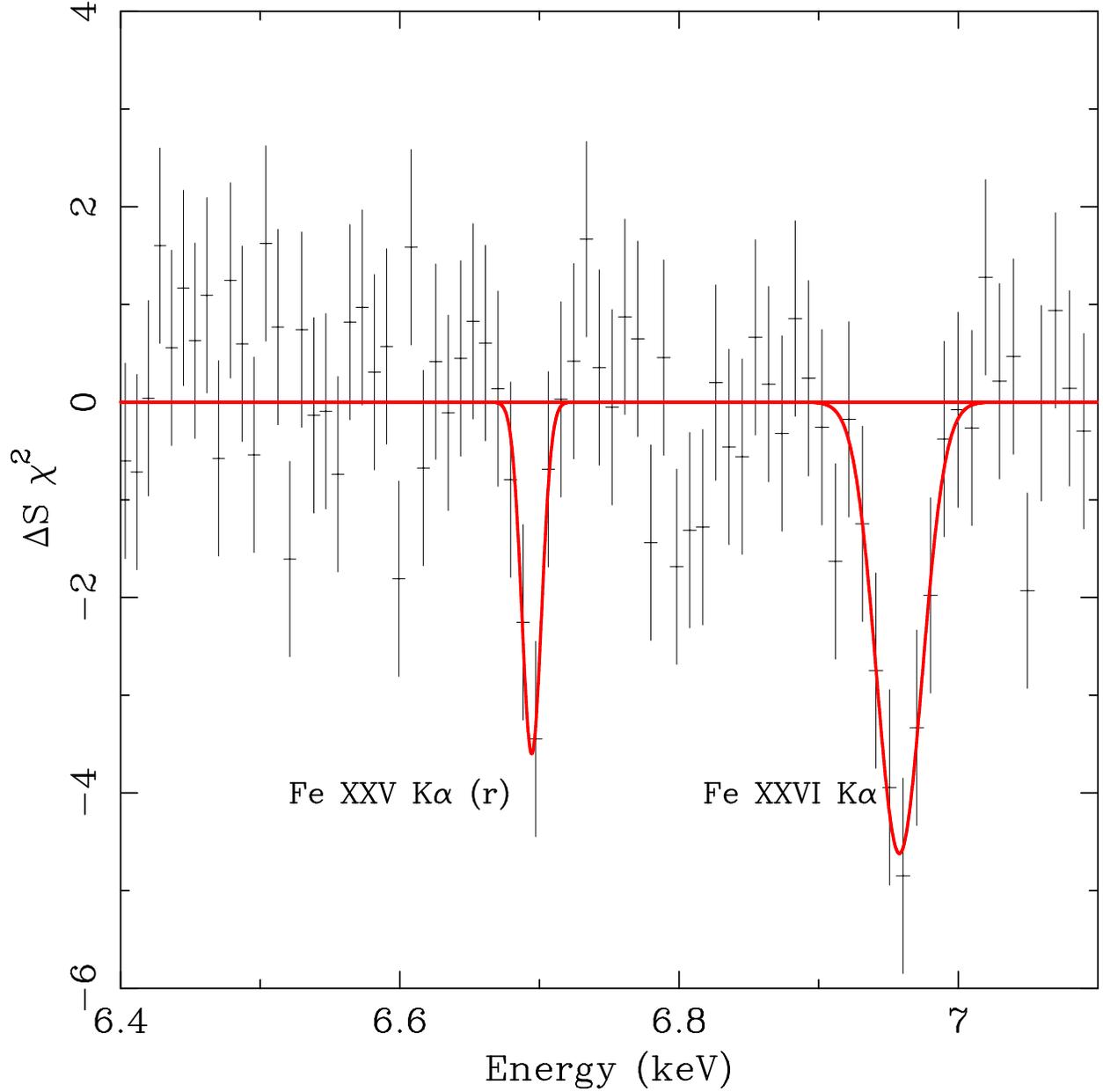}}
%\plotone{f1.ps}
\caption{Residuals of the 1st-order HEG spectrum, in the
  6.4--7.1 keV energy range, with respect to the best-fit model of the
  continuum  reported in  Tab.  \ref{tab1}.  Two  absorption features,
  identified as  \ion{Fe}{25} K$_\alpha$ and  \ion{Fe}{26} K$_\alpha$,
  are clearly observed}
\label{fig4}
\end{figure}

To resolve the absorption features  we fixed the continuum and added a
Gaussian line  with negative normalization for each  feature.  We used
the 1st-order MEG spectrum to  resolve the absorption features below 3
keV and  the 1st-order HEG spectrum  to resolve those in  the 6--7 keV
energy band.   We detected  four absorption lines  below 3  keV, these
were centered at  1.021, 1.471, 2.004, and 2.617  keV and corresponded
to  \ion{Ne}{10}  K$_\alpha$,  \ion{Mg}{12}  K$_\alpha$,  \ion{Si}{14}
K$_\alpha$, and  \ion{S}{16} K$_\alpha$, respectively;  the equivalent
widths  were -2.13,  -1.18,  -2.82, and  -3.56  eV, respectively.   In
Fig.~\ref{fig3} we  showed four expanded  views of the  residuals with
respect  to the  continuum  in  the narrow  energy  ranges around  the
centroids of  each of the absorbed  features.  In the  6--7 keV energy
range we  detected two  absorption lines centered  at 6.693  and 6.966
keV,  corresponding  to   \ion{Fe}{25}  K$_\alpha$,  and  \ion{Fe}{26}
K$_\alpha$;  the   equivalent  widths   were  -12.7,  and   -29.9  eV,
respectively.  In Fig.~\ref{fig4} we showed the residuals with respect
to the continuum in the 6.4--7.1 keV energy range, in Table~\ref{tab1}
we  reported the parameters  of the  continuum and  of each  line.  We
noted that the line energies did  not fit with the lab energies.  This
was not due to a physical  effect but to a systematic error associated
to the  uncertainty of 0.6\arcsec\  of the source position  which also
identifies  the zero-point  of the  dispersion arms.   Considering the
systematic error associated to  the absolute wavelength accuracy (i.e.
$\pm  0.006 {\rm  \AA}$ and  $\pm 0.011  {\rm \AA}$  for HEG  and MEG,
respectively) the line energies are compatible with the lab energies.

\section{Discussion}
We  have  analyzed a  42  ks  Chandra  observation of  the  persistent
emission from XB~1916--053. The  position of the zeroth-order image of
the source provides improved X-ray coordinates for XB~1916--053 (R.A.=
$19^h18^m47^s.871$, DEC=-05$^{\circ}$14\arcmin  17\arcsec.09), with an
angular separation of 8.7\arcsec\  to the optical counterpart (see Liu
et al., 2000)  and of 6\arcsec\ to the X-ray  position reported by the
``Coordinate  Converter''  tool (see  section  2)  .   We performed  a
spectral analysis  of the persistent emission using  the 1st-order MEG
and HEG spectra.  The continuum emission is well fitted by an absorbed
power  law with  photon  index 1.5.   The  equivalent hydrogen  column
density  of  the  absorbing  matter  was to  $  0.44  \times  10^{22}$
cm$^{-2}$, this  value is the same  obtained by Boirin  et al.  (2004)
analyzing the XMM RGS spectra  of XB~1916--053.  Also we note that the
unabsorbed luminosity,  in the 0.6--10  keV energy range, is  a factor
1.7  larger than  the previous  XMM observation,  this  conclusion was
furtherly  supported  by our  analysis  of  the  RXTE ASM  lightcurve.
Another interesting  detection is  that the dips  are not  observed at
each  orbital period,  this could  be explained  considering  that the
larger  flux from  the source  during our  observation  could largerly
photoionize the matter of the absorber at the disk edge.

We  clearly  detected  the  presence  of  the  \ion{Ne}{10}  (H-like),
\ion{Mg}{12}  (H-like),  \ion{Si}{14}(H-like),  \ion{S}{16}  (H-like),
\ion{Fe}{25}(He-like),  and  \ion{Fe}{26}  (H-like) absorption  lines.
The  \ion{Fe}{25}  and  \ion{Fe}{26}  absorption  lines  were  already
observed by Boirin  et al.  (2004) using the  XMM EPIC pn observation,
in that case the line widths had upper limits of $<100$ and $<140$ eV,
respectively. Thanks to the higher spectral resolution of Chandra HEG,
to an observation four times longer  and to a brightness of the source
two times larger, we found that  the line widths are $<13$ and between
0.1  and  21  eV  for  \ion{Fe}{25}  and  \ion{Fe}{26},  respectively.
Moreover Boirin et al.  (2004) marginally detected, in the RGS data, a
\ion{Mg}{12} absorption line with an  upper limit on the line width of
41 eV.   Also in this  case, thanks to  higher statistics, we  found a
more stringent upper  limit on the line width  of 1.3 eV.  Furthermore
we noted  that Boirin et  al.  (2004) detected  in the RGS  spectra an
absorption edge at $0.99 \pm  0.02$ keV during the persistent emission
of the source.  The larger statistics of our observation allowed us to
well fitted the absorption edge near 1 keV using an absorption line at
1.02 keV associated to \ion{Ne}{10}.  Finally we noted that the energy
of the  absorption edge was between  0.87 and 0.97 keV  in the dipping
energy spectra of  XB 1916--053 observed with XMM  RGS (Boirin et al.,
2004), we think that this discrete feature could be an absorption line
associated  to  \ion{Ne}{9},  this  possibility does  not  change  the
scenario proposed by Boirin et  al.  (2004) suggesting that during the
dips the absorbing matter is less photoionized.

Since  both  \ion{Fe}{25} and  \ion{Fe}{26}  absorption features  were
detected, some  physical parameters of the plasma  responsible for the
lines could  be estimated.   The column density  of each ion  could be
estimated from the EW of  the corresponding absorption line, using the
relation  quoted e.g.,  by  Lee  et al.   (2002,  see also  references
therein) linking  the two  quantities, which is  valid if the  line is
unsaturated and on  the linear part of the curve  of growth, which was
verified in the  case of XB 1916-053.  The  ratio between \ion{Fe}{25}
and \ion{Fe}{26} column  densities could then be used  to estimate the
photo-ionization parameter,  $\xi$, using the  calculations of Kallman
\&  Bautista  (2001).   Following  this approach,  we  derived  column
densities of  $1.5 \times 10^{17}$ cm$^{-2}$ and  $6.6 \times 10^{17}$
cm$^{-2}$ for  \ion{Fe}{25} and \ion{Fe}{26},  respectively.  We found
$\xi \sim  10^{4.15}$ erg cm  s$^{-1}$, that was slightly  larger than
the  ionization parameter  obtained from  XMM  data by  Boirin et  al.
(2004)  of $\xi_{XMM}  = 10^{3.92}$  erg cm  s$^{-1}$.  We  noted that
$\xi$  was a  factor  1.7  larger than  $\xi_{XMM}$,  the same  factor
obtained  comparing the unabsorbed  luminosity during  our observation
and  the  XMM  observation   (see  above).   Moreover  the  ionization
parameter associated to the H-like ions of Ne, Mg, Si, and S should be
$\sim 10^3$, that  was lower than that associated  to \ion{Fe}{25} and
\ion{Fe}{26} absorption lines.   It is worth to note  that both Boirin
et al.   (2004) and ourself  inferred the ionization  parameters $\xi$
basing on the  photoionized model by Kallman \&  Bautista (2001) which
assumed  an  ionization  continuum  consisting  of a  power  law  with
$\Gamma=1$, but Diaz Trigo et al.   (2005) have found a lower value of
the ionization  parameter of log$(\xi)=3.05 \pm 0.04$,  using a photon
index of $\Gamma=1.87 $ (obtained  fitting the XMM data), and assuming
a cutoff energy  of 80 keV as obtained by Church  et al.  (1998) using
BeppoSAX data.

\begin{figure}
\resizebox{\hsize}{!}{\includegraphics{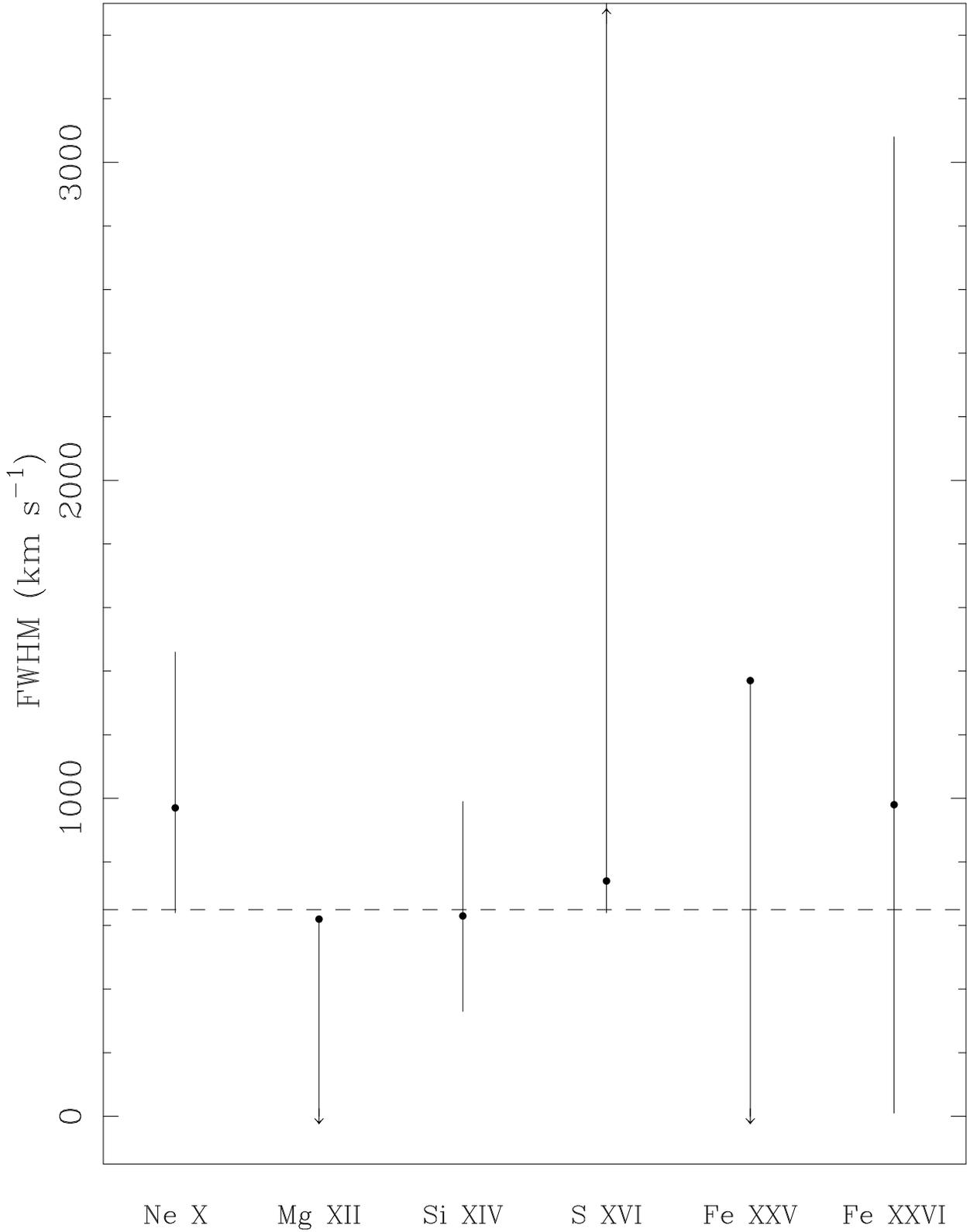}}
%\plotone{f1.ps}
\caption{ FWHM in km s$^{-1}$ of the observed absorption lines. All
the    FWHMs are compatible with a velocity of 650  km s$^{-1}$
indicated by the dashed horizontal line.}
\label{fig5}
\end{figure}

We  computed  the  FWHMs of  the  absorptions  lines  in units  of  km
s$^{-1}$,  these values  were  reported in  Table~\ref{tab1} and  were
plotted in  Fig.~\ref{fig5} in  which we note  that the values  of the
FWHMs  are compatible  with  a  velocity of  650  km s$^{-1}$  (dashed
horizontal  line).  We  investigated the  nature of  the  line widths.
Initially we  assumed that  the line widths  were produced  by thermal
broadening and we estimated  the plasma temperature using the relation
$kT= 511 m_I/m_e  (\sigma/E)^2$ keV, where $kT$ is  the temperature of
the plasma,  $m_I$ and  $m_e$ are  the mass of  the ion  and electron,
respectively, and $\sigma$  and $E$ are the width  and the centroid of
the  absorption  line  in   keV.   We  found  that  the  \ion{Ne}{10},
\ion{Mg}{12}, \ion{Si}{14}, and \ion{S}{16} absorption lines should be
produced in a region with a  temperature between 20 and 40 keV, while,
we found  an upper limit of $\sim  200$ keV to the  temperature of the
region where the \ion{Fe}{25}  and \ion{Fe}{26} absorption line should
be produced.   The interpretation of  the line as  thermally broadened
was not consistent with the  interpretation of the iron line ratios as
diagnostics of ionization parameter.  If the temperature was really 20
keV or  greater, then all the  elements would be  fully stripped, with
the possible exception of iron.  A more probable scenario was that the
line  widths  were  broadened   by  some  bulk  motion  or  supersonic
turbulence with a  velocity around to 650 km  s$^{-1}$ as indicated by
the FWHMs.   XB 1916--053  has an extended  corona around  the compact
object (Church  et al., 1998), assuming that  the mechanism generating
the turbulence or bulk motion was  due to the presence of the extended
corona  we can achieve  some informations  about where  the absorption
lines were produced.  Coronal models tend to have turbulent velocities
which are  locally proportional to  the virial or  rotational velocity
(Woods et al., 1996).  At $10^9$ cm (the coronal radius, see Narita et
al., 2003)  the virial  velocity is 4400  km/s, considering  a neutron
star of $1.4 M_{\odot}$.  It  is very difficult to construct plausible
dynamical  models in  which the  matter moves  at 10\%  of  the virial
velocity,  then  these  lines   should  be  produced  at  much  larger
distances, near  to $4 \times 10^{10}$  cm, i.e.  near  the disk edge.
According to this scenario we  concluded that the absorbing matter was
located at the same distance from  the neutron star of the bulge which
was  likely responsible for  the dips  themselves.  In  the hypothesis
that  the thickness  of the  absorbing region  was much  less  than $4
\times 10^{10}$ cm, its distance  from the source, we could estimate a
constraint on the  thickness $d$ of the absorber using $d  < L /\xi N$
(Reynolds \& Fabian 1995), where $L$ is the unabsorbed luminosity, $N$
is the equivalent hydrogen  column density of the photoionized matter,
and $\xi$  measures the corresponding  ionization parameter.  Assuming
the cosmic  abundance for  the iron and  a population  of \ion{Fe}{25}
with respect to neutral iron of 0.5 we found $d < 8 \times 10^4$ km.

\section{Conclusion} 

We  studied the  persistent emission  of XB  1916--053 using  a  42 ks
Chandra observation.  We improved the  position of the source, the new
coordinates        are       R.A.=        $19^h18^m47^s.871$       and
Dec.=-05$^{\circ}$14\arcmin 17\arcsec.09 (J2000.0) with an uncertainty
circle of the absolute position of 0.6\arcsec.

We   detected   the   \ion{Ne}{10},  \ion{Mg}{12},   \ion{Si}{14},   and
\ion{S}{16}  absorption lines  centered  at 1.021,  1.471, 2.004,  and
2.617 keV, respectively. These lines  were never observed before in XB
1916--053. Of all the  X-ray binaries exhibiting absorption lines only
Cir X--1  shows the same wide  series of absorption  lines, although in
that case P-Cygni profiles were evident (Brandt \& Schulz, 2000).

We  confirmed  the  presence   of  the  \ion{Fe}{25}  Ly$_\alpha$  and
\ion{Fe}{26} Ly$_\alpha$ absorption lines at 6.69 and 6.96 keV already
observed by XMM-Newton. From the study of equivalent widths of the two
lines we inferred a ionization parameter log($\xi$) of 4.15 that was a
factor  1.7 larger  than  during the  previous  XMM observation.   The
unabsorbed  luminosity in the  0.6--10 keV  energy range,  $7.5 \times
10^{36}$  erg s$^{-1}$,  was also  larger of  the same  factor  and we
verified this result studying the RXTE ASM lightcurve of XB 1916--053.
The increase of the ionization  parameter and of the luminosity of the
same quantity indicated that these two lines were produced in the same
region.   We  estimated  that  the  absorption line  widths  could  be
compatible  with  a  broadening  caused  by bulk  motion  or  turbulence
connected to the coronal activity,  finding from the broadening of the
absorption  lines that  these were  produced  at a  distance from  the
neutron star of $4 \times 10^{10}$ cm, i.e.  near the disk edge and at
the  same radius of  the absorber  which causes  the dipping  when the
corresponding ionization parameter decreases.

\acknowledgements We  are sincerely grateful to  the anonymous referee
for the useful  suggestions given to improve this  work.  This work was
partially  supported  by  the  Italian  Space  Agency  (ASI)  and  the
Ministero della Istruzione, della Universit\'a e della Ricerca (MIUR).

\clearpage

\begin{deluxetable}{lc}
\tabletypesize{\scriptsize}
%\tablewidth{8.5cm}
\tablecaption{Results of the spectral fit.}
\tablehead{\colhead{} &\colhead{} \\ 
\colhead{} &\colhead{Parameters}  }

\startdata
Continuum &  \\
\hline

$N_{\rm H}$ $\rm (\times 10^{22}\;cm^{-2})$ 
& $0.4448^{+0.0090}_{-0.0087}$ \\

photon index 
&  $1.4957^{+0.0099}_{-0.0095}$ \\  

N$_{po}$
&  $0.1108^{+0.0014}_{-0.0013}$ \\
&\\

\ion{Ne}{10} K$_\alpha$   &  \\
\hline

E$$ (keV)
& $1.02056^{+0.00073}_{-0.00043}$ \\

$\sigma$ (eV)
&  $1.40^{+0.71}_{-0.48}$ \\

I$$ ($\times 10^{-4}$ cm$^{-2}$ s$^{-1}$)
& $-2.24^{+0.55}_{-0.63}$ \\

EW$$ (eV)
&  $-2.13 \pm 0.57$ \\

FWHM$$ (km s$^{-1}$)
&  $970^{+490}_{-330}$ \\

&\\

\ion{Mg}{12}   K$_\alpha$ &  \\
\hline

E$$ (keV)
& $1.47116 ^{+0.00046}_{-0.00045}$  \\

$\sigma$ (eV)
&  $<1.3$ \\

I$$ ($\times 10^{-5}$ cm$^{-2}$ s$^{-1}$)
& $-7.4^{+1.8}_{-2.0}$ \\

EW$$ (eV)
&  $-1.18^{+0.29}_{-0.32}$ \\

FWHM$$ (km s$^{-1}$)
&  $< 620$ \\
&\\

\ion{Si}{14}  K$_\alpha$ &  \\
\hline

E$$  (keV)
& $2.00352  ^{+0.00071}_{-0.00072}$  \\

$\sigma$ (eV)
&  $1.79^{+1.03}_{-0.84}$ \\

I$$ ($\times 10^{-4}$ cm$^{-2}$ s$^{-1}$)
&  $-1.13^{+0.20}_{-0.21}$ \\

EW$$ (eV)
& $-2.82 \pm 0.53$ \\
FWHM$$ (km s$^{-1}$)
&  $630^{+360}_{-300}$ \\
&\\

\ion{S}{16} K$_\alpha$  &  \\
\hline

E$$  (keV)
& $2.61653^{+0.00472}_{-0.00082}$  \\

$\sigma$ (eV)
&  $2.76^{+158.03}_{-0.28}$ \\

I$$ ($\times 10^{-5}$ cm$^{-2}$ s$^{-1}$)
&  $-9.6^{+3.6}_{-1.9}$ \\

EW$$ (eV)
& $-3.56^{+1.32}_{-0.69}$  \\
FWHM$$ (km s$^{-1}$)
&  $740^{+42000}_{-100}$ \\
&\\

\ion{Fe}{25} K$_{\alpha}$  &  \\
\hline

E$$  (keV)
& $6.6925^{+0.0088}_{-0.0057}$  \\

$\sigma$ (eV)
&  $<13$   \\

I$$ ($\times 10^{-5}$ cm$^{-2}$ s$^{-1}$)
&  $-8.3^{+3.7}_{-3.5}$ \\

EW$$ (eV)
&  $-12.7 \pm 5.6$   \\
FWHM$$ (km s$^{-1}$)
&  $<1400$ \\
&\\
\ion{Fe}{26} K$_{\alpha}$   &  \\
\hline

E$$  (keV)
& $6.9558^{+0.0062}_{-0.0066}$  \\

$\sigma$ (eV)
&  $9.7^{+21.0}_{-9.6}$ \\

I$$ ($\times 10^{-4}$ cm$^{-2}$ s$^{-1}$)
&  $-1.83^{+0.46}_{-0.58}$ \\

EW$$ (eV)
& $-29.9^{+7.6}_{-9.4}$  \\
FWHM$$ (km s$^{-1}$)
&  $980^{+2100}_{-970}$ \\

\enddata  \tablecomments{The model  is  composed of  a power-law  with
  absorption   from  neutral  matter.    Uncertainties  are   at  90\%
  confidence level  for a single  parameter, upper limits are  at 95\%
  confidence  level.   N$_{po}$  indicates  the normalization  of  the
  power-law component in unit of photons keV$^{-1}$ s$^{-1}$ cm$^{-2}$
  at  1 keV.  The  parameters of  the Gaussian  emission lines  are E,
  $\sigma$, I,  and EW  indicating the centroid  in keV, the  width in
  keV,  the  intensity  of  the  line in  units  of  photons  s$^{-1}$
  cm$^{-2}$, and the equivalent width in eV, respectively. }
%%Our orbital solution includes the orbital period derivative; the reference 
%%epoch for $\Porb$ is given by ${\T0}$.}
\label{tab1}
\end{deluxetable}

\clearpage

\begin{thebibliography}{}

%\bibitem[]{}
%Bautista, M. A., \& Kallman, T. R., 2000, ApJ, 544, 581
\bibitem[]{}
Boirin, L., Parmar, A. N., Barret, et al.,  2004, A\&A, 418, 1061
\bibitem[]{}
Boirin, L., Mendez, M., Diaz Trigo, M., et al.,  2005, A\&A, 436, 195
\bibitem[]{}
Brandt, W. N., \& Schulz, N. S., 2000, ApJ, 544, L123
\bibitem[]{}
Callanan, P. J., Grindlay, J. E., \& Cool, A. M., 1995, PASJ, 47, 153
 \bibitem[]{}
 Church, M. J.  \& Balucinska-Church, M., 1995, A\&A, 300,  441
\bibitem[]{}
 Church, M.J.,  \&  Balucinska-Church, M., 2001, A\&A, 369,915
\bibitem[]{}
Church, M. J., Dotani, T., Balucinska-Church, M., et al., 1997, ApJ,
  491, 388
\bibitem[]{}
Church, M. J., Parmar, A. N.,  Balucinska-Church, M., et al., 1998,
A\&A, 338, 556
\bibitem[]{}
Davis, J. E. 2001, ApJ, 562, 575 
\bibitem[]{}
 Diaz Trigo, M.,  Parmar, A. N.,  Boirin , L., et al., 2005, ArXiv
 Astrophysics e-prints, astro-ph/0509342 
 \bibitem[]{}
Garmire, G. P., Bautz, M. W., Ford, et al., 2003, Proc. SPIE, 4851, 28 
\bibitem[]{}
Kallman, T., \& Bautista, M., 2001, ApJS, 133, 221
\bibitem[]{}
Kallman, T. R., \& McCray, R. 1982, ApJS, 50, 263
\bibitem[]{}
Kotani, T., Ebisawa, K., Dotani, T., et al., 2000, ApJ, 539, 413
\bibitem[]{}
Lee, J. C., Reynolds, C. S., Remillard, R., et al., 2002, ApJ, 567, 1102
\bibitem[]{}
Liu, Q. Z., van Paradjis, J., van den Heuvel, E. P. J., 2001, A\&A,
368, 1021  
\bibitem[]{}
Miller, J. M., Fabian, A. C., Wjinands, R., et al., 2002, 578, 348
\bibitem[]{}
Miller, J. M., Raymond, J., Homan, J., et al., 2004, ArXiv
 Astrophysics e-prints, astro-ph/0406272
\bibitem[]{}
Narita, T., Grindlay, J. E., Bloser, P. F., 2003, ApJ, 593, 1007
\bibitem[]{}
Parmar, A. N., White, N. E., Giommi, P.,  et al.,  1986, ApJ,
  308, 199
\bibitem[]{}
Retter, A., Chou, Y., Bedding, T. R., et al.,  2002, MNRAS, 330,  L37
\bibitem[]{}
Reynolds, C. S., \& Fabian, A. C., 1995, MNRAS, 273, 1167
\bibitem[]{}
Ueda, Y., Inoue, H., Tanaka, Y., et al., 1998, ApJ, 492, 782
\bibitem[]{}
Ueda, Y., Murakami, H., Yamaoka, K., et al.,  2004, ApJ, 609, 325
\bibitem[]{}
Walter, F. M., Mason, K. O., Clarke, J. T., et al., 1982, ApJ, 253, L67
\bibitem[]{}
White, N. E., \& Holt, S. S., 1982, ApJ, 257, 318
\bibitem[]{}
White, N. E., \& Swank, J. H. 1982, ApJ, 253, L61
\bibitem[]{}
Woods, D. T.,  Klein, R. I., Castor, J. I., et al., 1996, ApJ, 461, 767 
\bibitem[]{} 
Yamaoka, K., Ueda, Y., Inoue, H., et al., 2001, PASJ, 53, 179
\bibitem[]{} 
Yoshida, K., Inoue,  H., Mitsuda, K., et al., 1995, PASJ, 47, 14



\end{thebibliography}
\end{document}